\documentclass[apjl, a4]{emulateapj}
\usepackage{natbib}
\usepackage{xspace}
\usepackage{graphicx}
\usepackage{hyperref}
\usepackage[normalem]{ulem}
\usepackage{amsmath}
\usepackage{amssymb}
\usepackage{ulem}
\usepackage{amsbsy}
\usepackage{mathrsfs,bm}
\usepackage{soul}
\newcommand{\bcdot}{\ensuremath{%
  \mathchoice%
   {\mskip\thinmuskip\lower0.2ex\hbox{\scalebox{1.5}{$\cdot$}}\mskip\thinmuskip}}%
   {\mskip\thinmuskip\lower0.2ex\hbox{\scalebox{1.5}{$\cdot$}}\mskip\thinmuskip}%
   {\lower0.3ex\hbox{\scalebox{1.2}{$\cdot$}}}%
   {\lower0.3ex\hbox{\scalebox{1.2}{$\cdot$}}}%
}
\usepackage[usenames]{color}

\newcommand{\bnabla}{\ensuremath{\boldsymbol{\nabla}}}
\newcommand{\vect}[1]{\boldsymbol{#1}}

\usepackage{xcolor}
\hypersetup{
    colorlinks,
    linkcolor={red!50!red},
    citecolor={blue!50!blue},
    urlcolor={blue!80!black}
}

\def\del#1{{}}

\newcommand{\rmn}{\mathrm}

\newcommand\scalingFirst{1.}
\newcommand\scalingSecond{0.49}
\shorttitle{The Sunyaev-Zel'dovich effect of simulated jet-inflated bubbles in clusters}
\shortauthors{Ehlert et al.}

\begin{document}

\title{The Sunyaev-Zel'dovich effect of simulated jet-inflated bubbles in clusters}

\author{Kristian Ehlert\altaffilmark{1}, Christoph Pfrommer\altaffilmark{1}, Rainer Weinberger\altaffilmark{2}, R{\"u}diger Pakmor\altaffilmark{3} and Volker Springel\altaffilmark{3}}
\email{kehlert@aip.de}

\altaffiltext{1}{Leibniz-Institut f{\"u}r Astrophysik Potsdam (AIP), An der Sternwarte 16, 14482 Potsdam, Germany}
\altaffiltext{2}{Center for Astrophysics $|$ Harvard \& Smithsonian, 60 Garden Street, Cambridge, MA 02138, USA}
\altaffiltext{3}{Max-Planck-Institut f\"ur Astrophysik, Karl-Schwarzschild-Str. 1, D-85741 Garching, Germany}

\begin{abstract}
Feedback by active galactic nuclei (AGNs) is essential for regulating the fast radiative cooling of low-entropy gas at the centers of galaxy clusters and for reducing star formation rates of central ellipticals. The details of self-regulation depend critically on the unknown contents of AGN-inflated bubbles. Observations of the Sunyaev-Zel’dovich (SZ) signal of AGN bubbles provide us with the ability to directly measure the lobe electron pressure given a bubble morphology. Here we compute the SZ signal of jet-inflated bubbles in three-dimensional magnetohydrodynamical simulations of the galaxy cluster MS0735.6+7421 with the Arepo code, and compare our synthetic SZ results to inferences obtained with popular modelling approaches.  We find that cutting out ellipsoidal bubbles from a double-beta pressure profile only matches the inner bubble edges in the simulations and fails to account for the emission of the shock-enhanced pressure cocoon outside the bubbles. This additional contribution significantly worsens the accuracy of the cut-out method for jets with small inclinations with respect to the line of sight. Also, the kinetic SZ effect of the bubbles, a previously neglected contribution, becomes relevant at these smaller inclinations due to entrainment and mixing of the intracluster medium with low-density jet material. Fortunately, the different signs of the kinetic SZ signal in opposite lobes allow modelling this effect. We present an approximate method to determine the jet inclination, which combines jet power and lifetime estimates, the stand-off distance between jet head and bow shock, and the kinetic SZ effect, thereby helping to correctly infer the bubble contents.
\end{abstract}

\keywords{galaxies: clusters: intracluster medium --- galaxies: clusters: individual (MS0735.6+7421) ---
  galaxies: active --- cosmic background radiation --- cosmic rays  --- magnetohydrodynamics (MHD)}

\section{Introduction}
\label{sec:intro}
The radiative cooling time of the intra-cluster medium (ICM) in the center of cool core clusters is less than $1~\mathrm{Gyr}$. The central AGN provides a powerful heating source that offsets cooling and reduces star formation \citep{McNamara2012}. AGN jets power lobes, which detach and buoyantly rise in the cluster atmosphere. However, the detailed lobe content and thus, the heating mechanism remains uncertain. While X-ray observations can only provide lower limits to the temperature of lobes \citep{Worrall2009}, the SZ signal \citep{Sunyaev1972} is directly sensitive to the thermal and non-thermal heat contents. Hence, SZ observations of bubbles have been suggested \citep{Pfrommer2005} and simulated \citep{Prokhorov2012,Yang2018} to understand the physics of the heating mechanism. Recently, \cite{Abdulla2018} observed the cavities of the cluster MS 0735.6+7421 (hereafter MS0735) and found that they have very little SZ-contributing material. This suggests a lobe pressure support of diffuse thermal plasma with temperature in excess of hundreds of keV, or non-thermal relativistic particle populations.

Assuming energy equipartition between relativistic electrons and magnetic fields, radio synchrotron observations also necessitate an additional pressure component \citep{Birzan2008}, suggesting relativistic protons as a likely candidate that matches jet morphologies \citep{Croston2018}. As these cosmic ray (CR) protons escape into the ICM, they resonantly excite Alfv\'en waves. Damping of those waves provides a promising heating scenario \citep{Guo2008,Ensslin2011,Fujita2012,Pfrommer2013,Jacob2016a,Jacob2016b}. Jet-driven bubble simulations in a galaxy cluster \citep{Ruszkowski2017a,Ehlert2018} demonstrate that streaming CRs from the bubbles provide a sufficient heating rate to halt the cooling catastrophe. Alternative AGN heating mechanisms include mixing of hot-bubble gas with the ICM \citep{Soker2016}, dissipation of sound waves \citep{Fabian2017}, weak shocks \citep{Li2016a}, turbulence \citep{Zhuravleva2014a,Bambic2018a} and/or gravity waves \citep{Bambic2018}. 

Early simulations of AGN bubbles reproduce the main features of observed X-ray cavities \citep{Churazov2001, Bruggen2009}. Simulations including magnetic fields \citep{Robinson2004,Ruszkowski2007} and/or viscosity \citep{Reynolds2005,Sijacki2006} stabilize the bubble against developing fluid instabilities. Turbulence and substructure show a significant impact on bubble dynamics \citep{Heinz2006,ONeill2010,Mendygral2012}. The addition of CRs in bubbles leads to more elongated bubbles that reside closer to the cluster center, which is favoured by observations \citep{Sijacki2008,Guo2011}. Recent simulations gained higher resilience against numerical mixing due to sophisticated refinement criteria \citep{Weinberger2017, Bourne2017}. This allows for more realistic (higher) density contrasts between simulated bubbles and ICM on long time scales.

For the first time, we use MHD jet simulations to study how the thermal and kinetic SZ signal of dynamical jet-blown bubble simulations compare to the simplified modelling of bubbles as ellipsoids. To facilitate observational comparison, we simulate CR-filled bubbles in a turbulent and magnetized cluster. We pick the observationally favoured, largest observed AGN outbreak in MS0735 to exemplify our analysis \citep{McNamara2005,Colafrancesco2005}.

We describe our simulation methods and setup in Section \ref{sec:simulations}. In Section \ref{sec:evolvingSZfromsimulatedbubbles}, we summarize the characteristics of jet evolution and the details of our SZ modelling. We show the expected signal of simulated bubbles with different fillings in Section \ref{sec:ProbingRelativisticFillings} and compare the SZ signal from a simulated bubble to a modelled ellipsoidal bubble. In Section \ref{sec:DegeneraciesInclination}, we discus our findings regarding the kinetic SZ effect from our simulated bubbles and the influence of jet inclination on the SZ signal. We propose a method of combining simulations and observations to constrain jet inclination enabling more stringent limits on bubble content. We conclude in Section~\ref{sec:conclusion}.

\section{Simulations}
\label{sec:simulations}

To study the SZ signal from an AGN bubble, we simulate a jet, which self-consistently inflates an MS0735-like bubble in a turbulently magnetized cluster atmosphere. We use the same simulation techniques as for the fiducial run in \cite{Ehlert2018} with parameters adopted to the outburst and ICM in MS0735.

The dark matter profile is modelled after MS0735 as a static Navarro-Frenk-White (NFW) profile with $M_{200,\mathrm{c}}=1.5\times10^{15}~\mathrm{M}_\odot$, $R_{200,\mathrm{c}}=2.43~\mathrm{Mpc}$ and concentration parameter $c_\mathrm{NFW}=3.8$ \citep{Gitti2007}. The electron number density follows a double-beta profile fit to MS0735 \citep{Vantyghem2014a} modified to obtain a gas fraction of 16\% at $R_{200,\mathrm{c}}$:
\begin{equation}
\begin{split}
n_\mathrm{e}&=0.05\left[1+\left(\frac{r}{100\ \text{kpc}}\right)^2\right]^{-4.9}\text{cm}^{-3}\\
 &\quad +0.01\left[1+\left(\frac{r}{400~\text{kpc}}\right)^2\right]^{-1.6}\text{cm}^{-3}.
\end{split}
\end{equation}
The Gaussian-distributed, turbulent magnetic field is generated in Fourier space and exhibits a Kolmogorov power spectrum on scales larger than the injection scale $k_\mathrm{inj}=37.5^{-1}~\mathrm{kpc}^{-1}$. On scales $k<k_\mathrm{inj}$ the spectrum follows a white noise distribution. The magnetic field is scaled in concentric spherical shells to obtain a predefined average magnetic-to-thermal pressure ratio of $X_{B,\mathrm{ICM}}=0.05$. Multiple nested meshes of magnetic fields with decreasing resolution from the central AGN, respectively, are required for the large range in spatial resolution within the box. Overlapping regions of neighbouring meshes are iteratively smoothed and cleaned off magnetic divergence. The initial Cartesian mesh is relaxed to obtain a honeycomb-like structure, which is more efficient for the unstructured, moving mesh code \textsc{arepo}. When evolved, the magnetic field drives turbulence through tension forces, which gradually decrease the magnetic field strength. Thus, we rescale the magnetic field to the desired $X_{B,\mathrm{ICM}}$ to obtain our initial conditions \citep[see][for further details]{Ehlert2018}.

We model the jets by injecting kinetic energy in two spherical regions with radius $r_j=1.65\ \mathrm{kpc}$ on opposite sides of the centrally placed supermassive black hole (BH) particle. Mass is removed from these injection regions and thermal energy is added from the surroundings to obtain low-density jets ($\rho_\mathrm{jet}=10^{-28}\ \mathrm{g}\ \mathrm{cm}^{-3}$, $\rho_\mathrm{jet}/\rho_\mathrm{ICM}\sim10^{-4}$) in pressure equilibrium with the ICM \citep[for more details, see][]{Weinberger2017}. Our fiducial run features a jet with power $P_\mathrm{jet}=2\times10^{46}\ \mathrm{erg}~\mathrm{s}^{-1}$ and lifetime $\tau_\mathrm{jet}=150~\mathrm{Myr}$, amounting to an injected energy $E_\mathrm{jet}=P_\mathrm{jet}\tau_\mathrm{jet}\approx10^{62}~\mathrm{erg}$ in MS0735 \citep{Vantyghem2014a}. In addition, we inject a helical magnetic field in the jet region with magnetic-to-thermal pressure ratio $X_{B,\mathrm{jet}}=0.1$. Our lobes are defined via an advective scalar $X_\mathrm{jet}$, which is set to unity in the jet injection region. We define our lobes as the union of all cells with $X_\mathrm{jet}>10^{-3}$. During jet activity CRs are accelerated in the lobes by converting thermal energy to CR energy whenever the CR-to-thermal pressure $X_\mathrm{cr}$ ratio falls below a predefined value $X_\mathrm{cr}<X_\mathrm{cr,acc}=1$. We explicitly isolate our jet injection region magnetically to inhibit unphysical CR diffusion.

In addition to advection, CRs are expected to scatter on self-excited Alfv\'en waves in galaxy clusters \citep{Kulsrud1969,Ensslin2011}. The low efficiency of Alfv\'en wave damping in the ICM limits the CRs to \emph{stream} down their pressure gradient $\bnabla {P}_\mathrm{cr}$ along magnetic field lines at the Alfv\'en speed $\vect{v}_\mathrm{A}$ \citep{Wiener2013}. The damping of Alfv\'en waves effectively transfers CR to thermal energy, giving rise to the so-called Alfv\'en heating with a power $\mathcal{H}_\mathrm{cr}=\left|\vect{v}_\mathrm{A}\bcdot \bnabla {P}_\mathrm{cr}\right|$.

The equations of MHD are discretized on a moving-mesh and evolved with second-order accuracy using the massively parallel \textsc{arepo} code \citep{Springel2010,Pakmor2016}. Cosmic rays are treated as a second fluid including hadronic and Coulomb losses \citep{Pfrommer2017}. These losses are small in comparison to Alfv\'enic losses, which we include here. In combination with anisotropic diffusion \citep{Pakmor2016a} (with a parallel diffusion coefficient $\kappa_{\parallel}=10^{29}\ \mathrm{cm}^{2}\ \mathrm{s}^{-1}$) this is used to emulate CR streaming. We employ mass-based refinement with target mass $m_\mathrm{target}=1.5\times10^{6}\ M_\odot$. In addition, we impose a refinement criterion based on the density gradient, jet scalar and cell volume difference as in \cite{Weinberger2017} to maintain the large density contrast at the jet-ICM interface with target volume $V_\mathrm{target}$ ($V^{1/3}_\mathrm{target}=405\ \mathrm{pc}$). 

\section{Evolving SZ signal from simulated bubbles}
\label{sec:evolvingSZfromsimulatedbubbles}
\begin{figure*}
\centering
\includegraphics[trim=2.2cm 0.4cm 2.4cm 1.6cm,clip=true,width=\scalingFirst\textwidth]{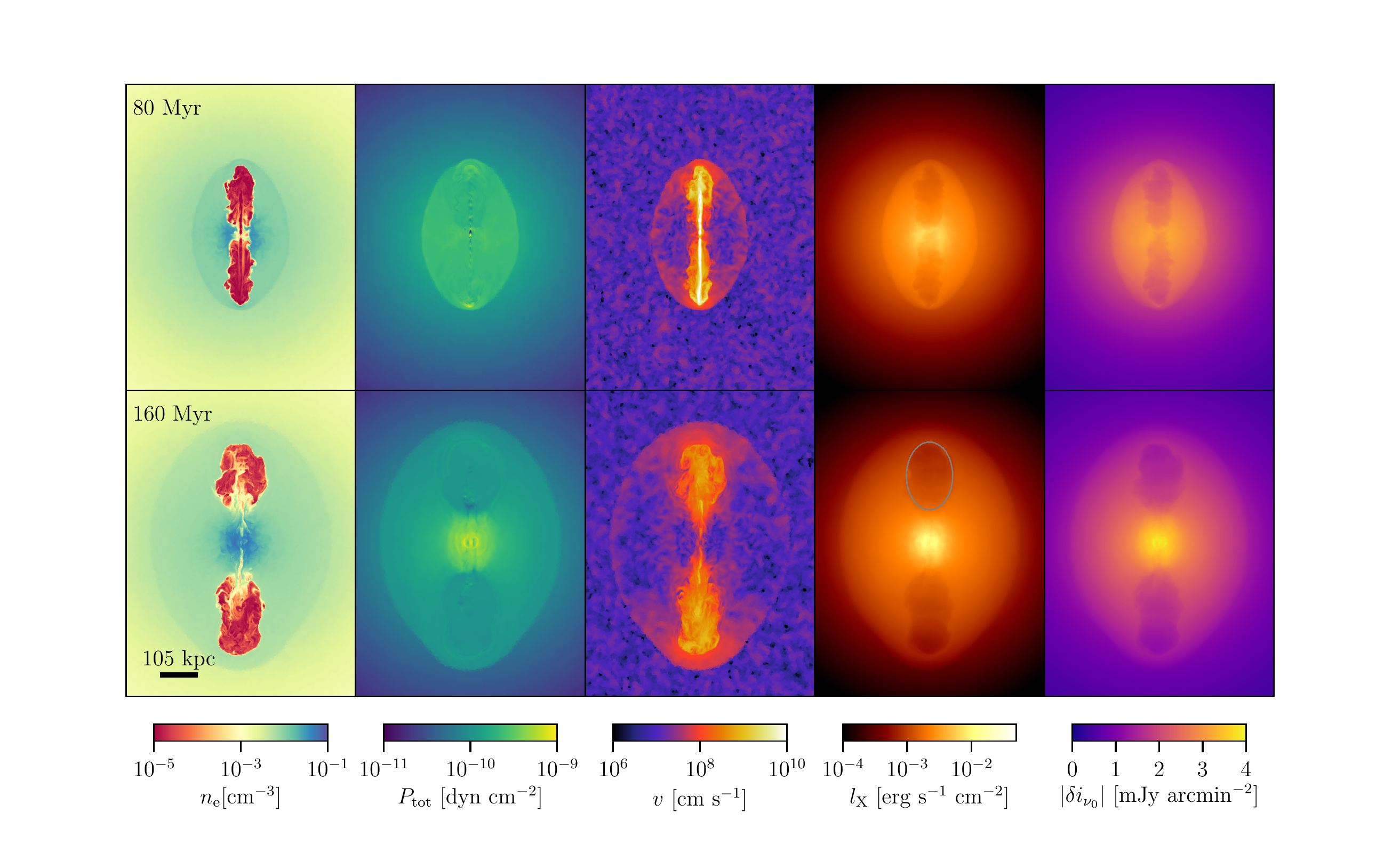}
    \caption{We show the electron density $n_e$, total pressure $P_\mathrm{tot}$
      ($P_\mathrm{tot}=P_\mathrm{th}+P_\mathrm{cr}+P_B$), velocity $v$, X-ray
      emissivity integrated along the line-of-sight $l_X$ and the SZ signal
      $|\delta i_{\nu_o}|$ of the fiducial run at times $80$ Myr and $160$ Myr
      (top and bottom row, respectively). For the SZ effect, we assume a bubble
      filled with relativistic electrons and an observing frequency
      $\nu_0=30\ \mathrm{GHz}$. The images correspond to projections of thin
      layers ($1000\ \mathrm{kpc}\times750\ \mathrm{kpc}\times4\ \mathrm{kpc}$)
      centred on the BH and weighted by cell volume except for the
      density-weighted velocity. The jet terminates at 150 Myrs after which
      buoyantly rising bubbles form that can be observed as cavities in X-ray
      and SZ images. The grey contour exemplifies our ellipsoidal bubble
        model.}
    \label{fig:evolution_absolute}
\end{figure*}

\begin{figure*}
\centering
\includegraphics[trim=0cm 0cm 0cm 0cm,clip=true, width=\scalingSecond\textwidth]{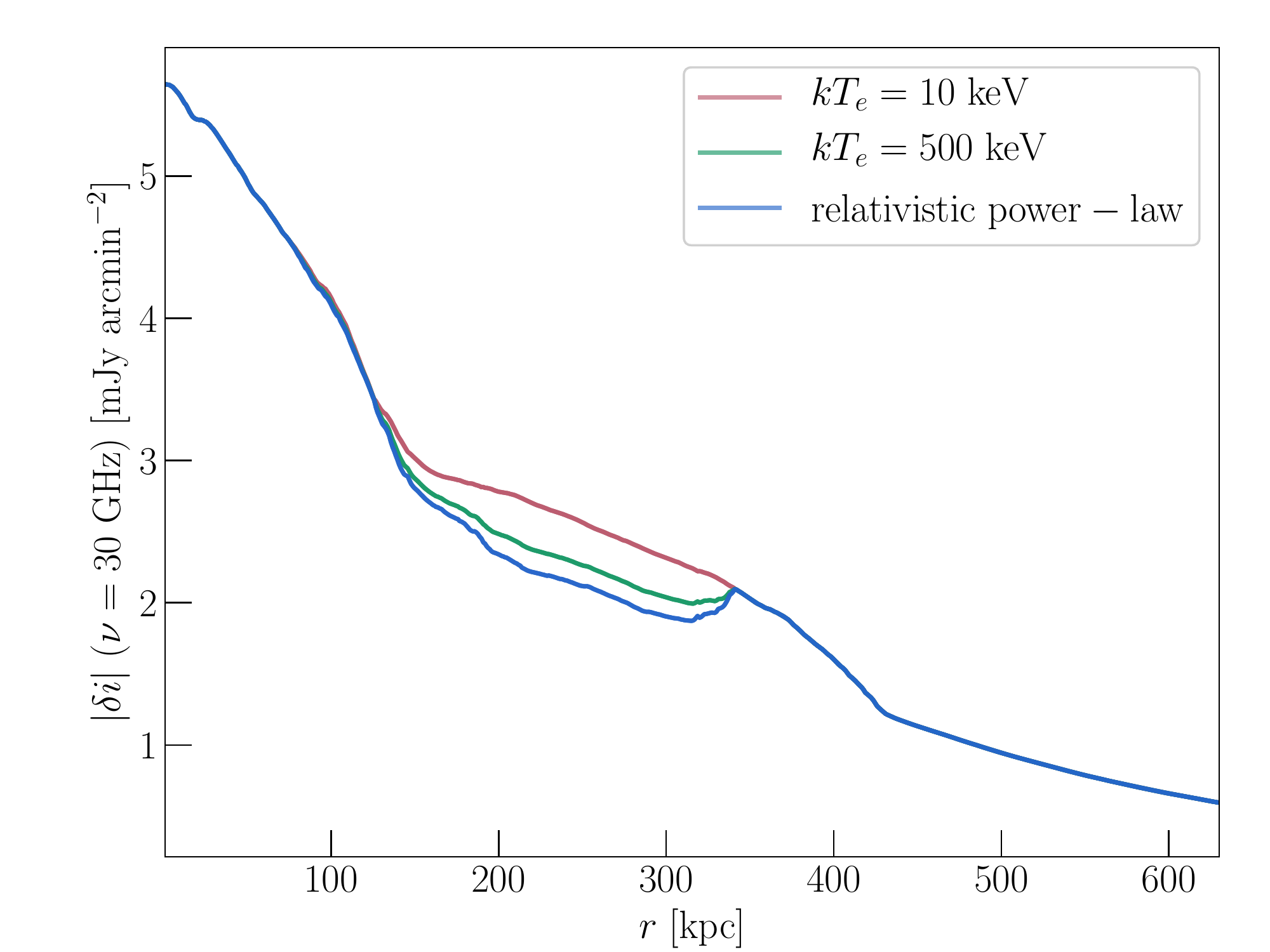}
\includegraphics[trim=0cm 0cm 0cm 0cm,clip=true, width=\scalingSecond\textwidth]{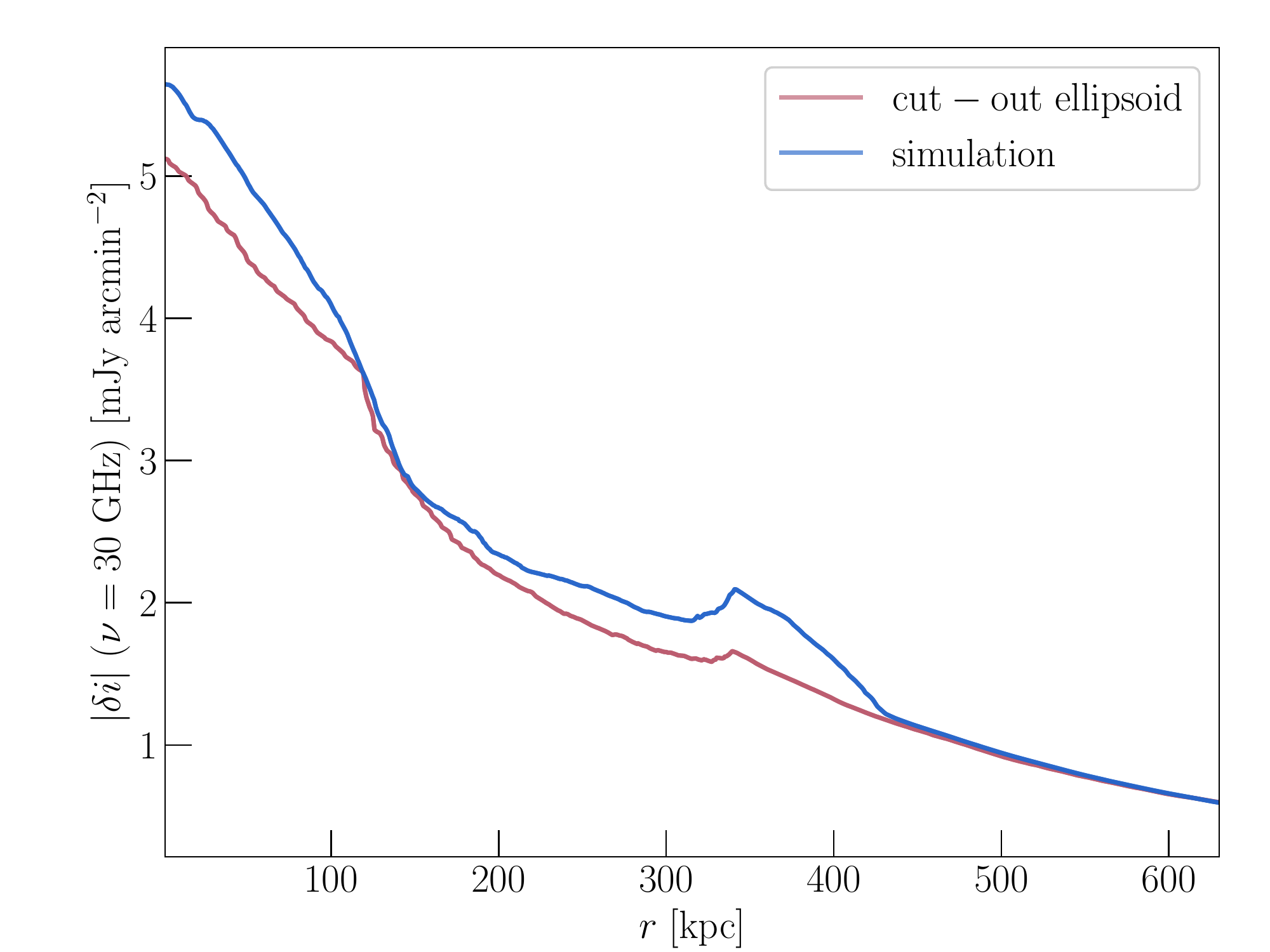}
\caption{On the left-hand side, we show profiles of the SZ signal
  along the jet axis at $180\ \mathrm{Myr}$. We vary the filling from
  thermal ($kT_\rmn{e}=10\ \mathrm{keV}$) electrons to relativistic
  electrons that either follow a thermal distribution
  ($kT_\rmn{e}=500\ \mathrm{keV}$) or a power-law
  distribution. Relativistically filled bubbles show a significantly
  larger SZ contrast, which is used to constrain the bubbles'
  content. On the right-hand side, we contrast the profiles of our
  simulated bubble to an ellipsoidal cut-out from our initial
  conditions, in which the jet outburst is viewed perpendicular to the
  line of sight. The cut-out approximation mimics the signal of the
  simulated bubble well at the inner rim and starts to deviate towards
  the bubble edges due to the increased pressure of the shocked
  cocoon that surrounds the bubble in the simulation. }
  \label{fig:radial_sz}
\end{figure*}

\begin{figure*}
\centering
\includegraphics[trim=2.79cm 2.135cm 0.1cm 0cm,clip=true, width=0.96\textwidth]{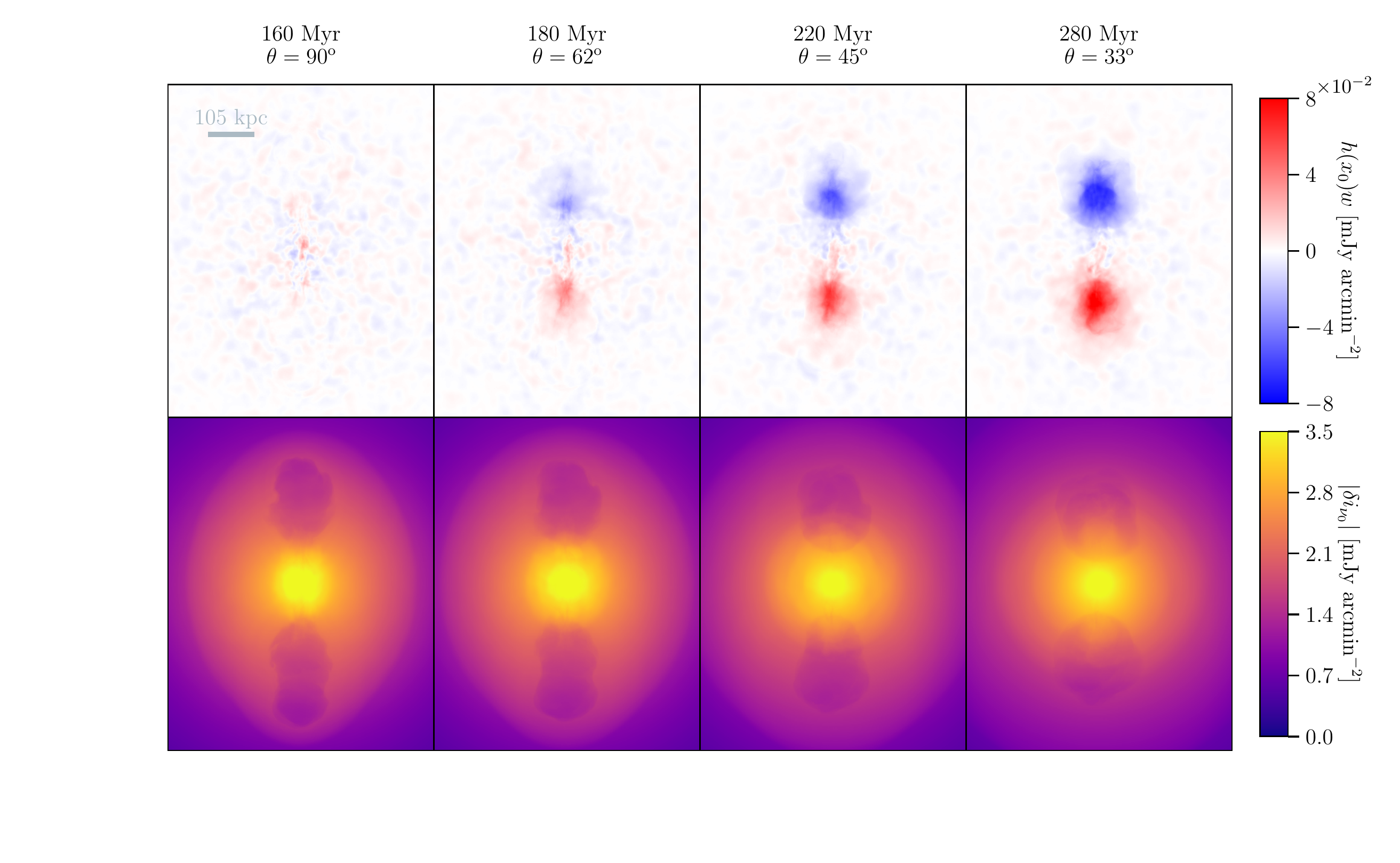}
\hspace*{-0.0255\textwidth}
\includegraphics[trim=0.2cm .3cm .75cm 1.273cm,clip=true, height=0.229\textheight]{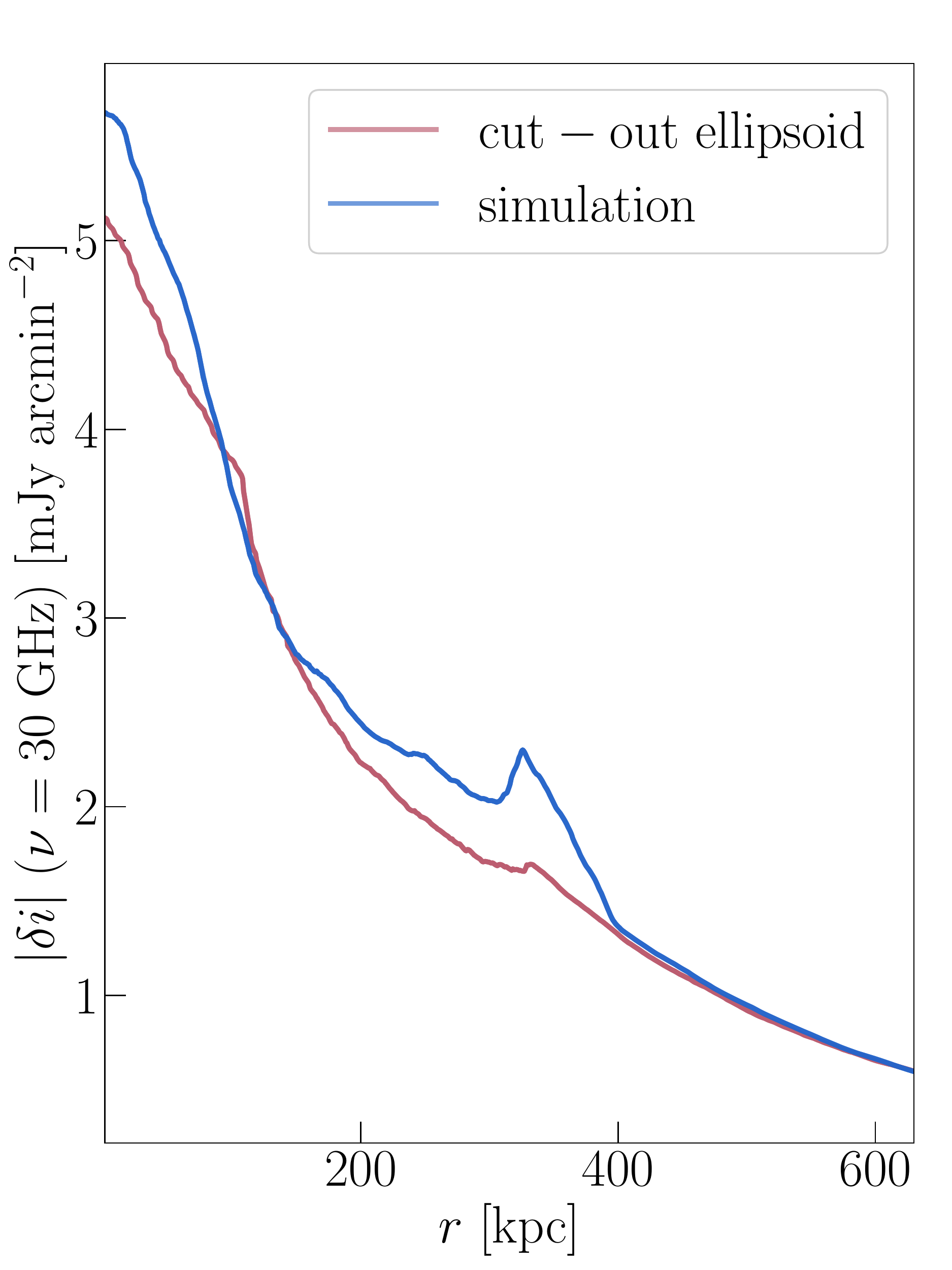}
\hspace{-0.0114\textwidth}
\includegraphics[trim=2.08cm .3cm .75cm 1.273cm,clip=true, height=0.229\textheight]{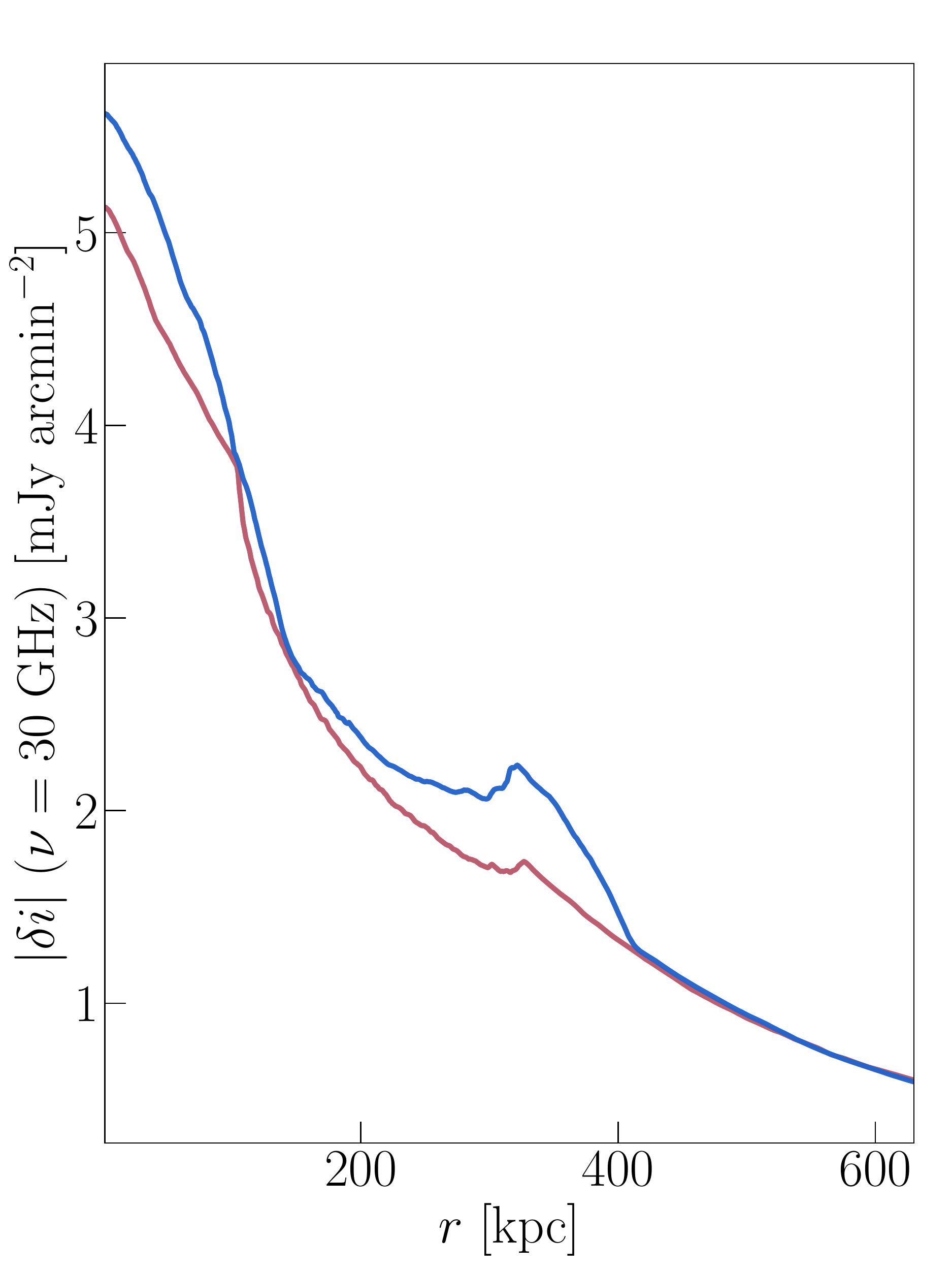}
\hspace{-0.0114\textwidth}
\includegraphics[trim=2.08cm .3cm .75cm 1.273cm,clip=true, height=0.229\textheight]{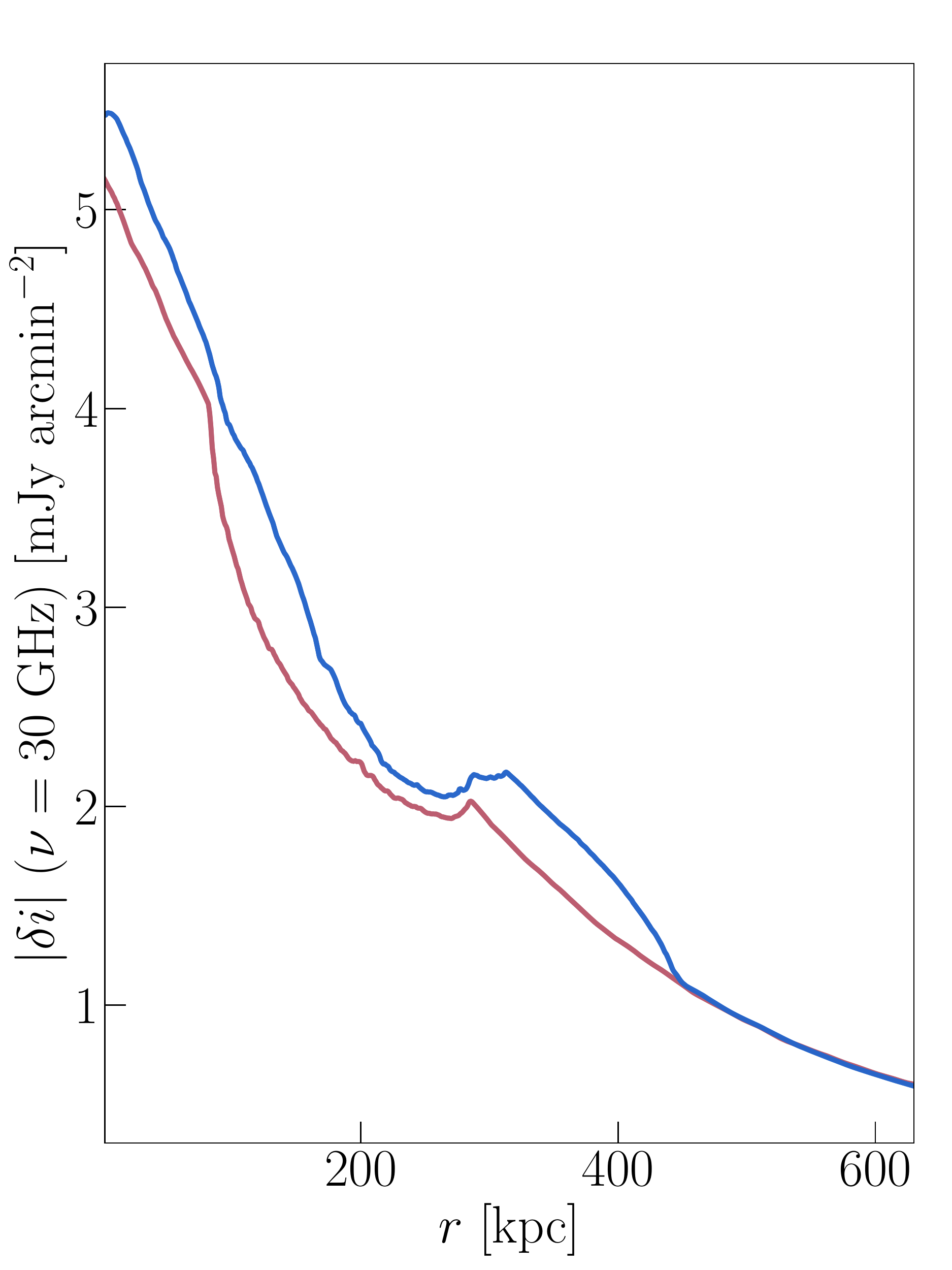}
\hspace{-0.0114\textwidth}
\includegraphics[trim=2.08cm .3cm -8.2cm 1.273cm,clip=true, height=0.229\textheight]{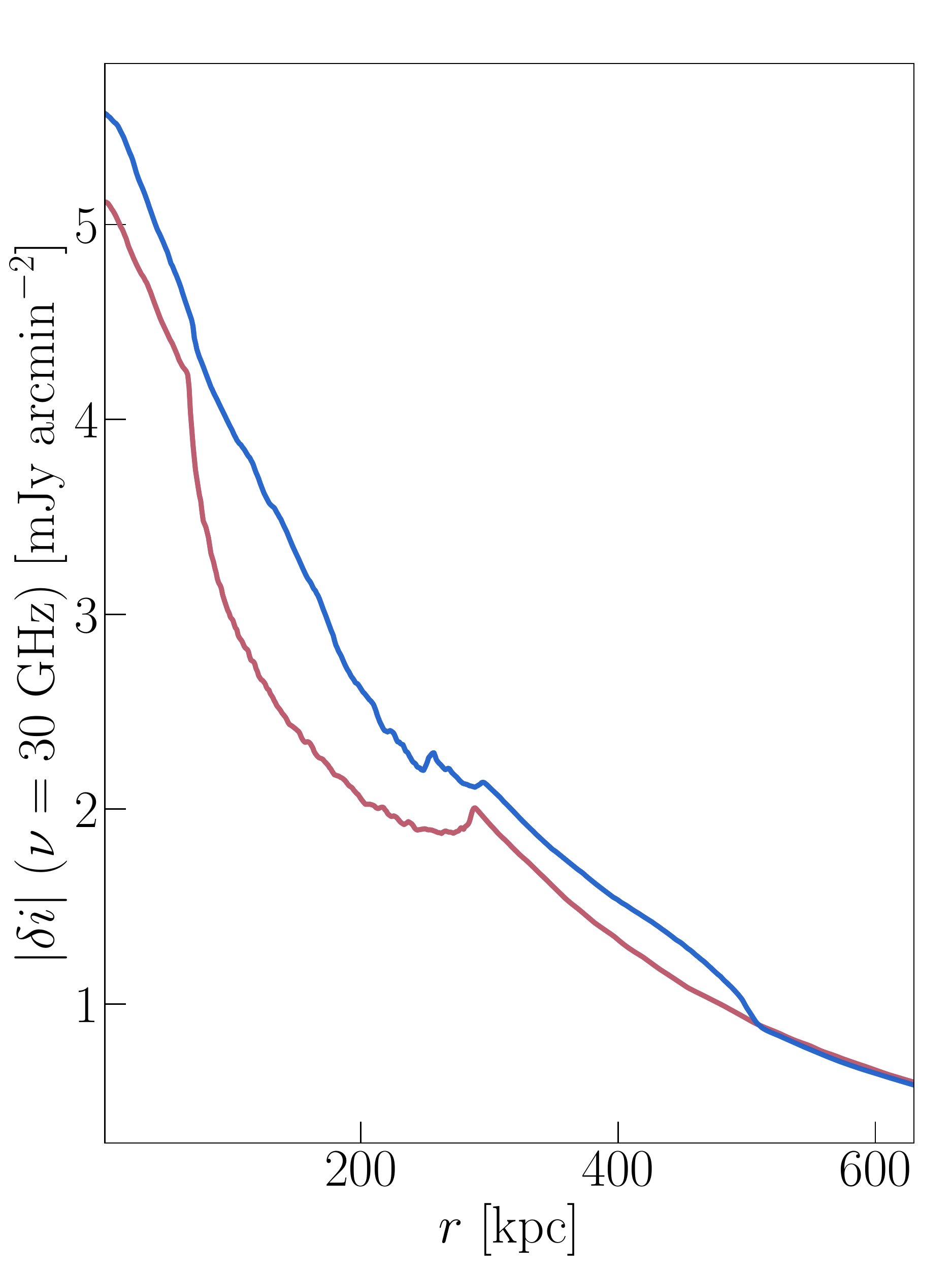}
\caption{The top row shows the integrated kinetic SZ effect $h(x_0)w$ at
  frequency $\nu_0=30\ \mathrm{GHz}$. The panels in the middle row show the
  thermal SZ signal for bubbles, which are filled with a relativistic power-law
  electron distribution. The angle of inclination $\theta$ between jet axis and
  line of sight is decreased from left to right. The images have dimensions
  $1000\ \mathrm{kpc}\times750\ \mathrm{kpc}$ and are centred on the BH. At low
  angles, i.e., $\theta=33^\circ$, the kinetic SZ signal can reach values of up
  to 10\% of the total SZ signal. The lower panel shows profiles of the SZ
  signal along the jet axis of the simulated bubbles (blue) and the
  cut-out model (red). While the model matches the simulations at the inner
  bubble edge for high values of inclinations, it differs significantly for
  lower inclinations as CMB photons intersect a larger portion of the
  ellipsoidal shocked cocoon including the central dense cool core region.}
    \label{fig:inclination_comparison}
\end{figure*}	

Free electrons in the ICM Compton up-scatter cosmic microwave background (CMB) photons. We follow the procedure described in \cite{Pfrommer2005} to compute the resulting SZ signal of our ICM and bubbles. The relative change $\delta i(x)$ in the flux density as a function of dimensionless frequency $x=h\nu/(k T_\mathrm{CMB})$ is given by
\begin{equation}
\label{eq:SZ}
\begin{split}
\delta i(x)=&g(x)y_\mathrm{gas}[1+\delta(x,T_e)]-h(x)w_\mathrm{gas}\\
&+[j(x)-i(x)]\tau_\mathrm{rel}.
\end{split}
\end{equation} 
with the Planckian distribution of the CMB given by
\begin{equation}
I(x)=i_0 i(x)=i_0\frac{x^3}{\exp(x)-1},
\end{equation}
where $i_0=2(kT_\mathrm{CMB})^3/(hc)^2$ and $T_\mathrm{CMB}=2.725\ \mathrm{K}$.
The first term in Equation \eqref{eq:SZ} describes the thermal SZ effect that is proportional to the integrated thermal pressure of the ICM along the line of sight $y_\mathrm{gas}\propto\int dl\ n_\mathrm{e,gas}kT_\mathrm{e}$. Relativistic corrections $\delta(x,T_\mathrm{e})$ become relevant at high temperatures $kT_e\gtrsim 5\ \mathrm{keV}$ \citep{Mroczkowski2018}. Throughout the analysis, we include relativistic corrections from \cite{Itoh1998}. The second term corresponds to the kinetic SZ effect due to gas motion relative to the Hubble flow $w_\mathrm{gas}\propto\int dl\ n_\mathrm{e,gas}v_\mathrm{gas}$, where $v_\mathrm{gas}<0$ if the gas is approaching the observer. The last term describes the relativistic SZ effect. For fully relativistic fillings the distribution function of the electrons determines the form of $j(x)$ and $\tau_\mathrm{rel}\propto \int dl\ n_\mathrm{e,rel}$. Only the terms $g(x)$, $h(x)$ and $i(x)$ depend on observing frequency.

Throughout the Letter, we focus on three different bubble fillings: 1.) a thermal electron distribution with $kT_\mathrm{e}=10\ \mathrm{keV}$, 2.) a single-temperature, relativistic Maxwellian with $kT_\mathrm{e}=500\ \mathrm{keV}$ and 3.) a single-temperature, relativistic electron population that follows a power-law in normalized momentum-space $p=\gamma_\mathrm{e}\beta_\mathrm{e}$ defined by
\begin{equation}
f_{\mathrm{CRe}}(p,\alpha,p_1,p_2)=\frac{(\alpha-1)p^{-\alpha}}{p_1^{1-	\alpha}-p_2^{1-\alpha}},
\end{equation}
where $\alpha=2$, $p_1=1$ and $p_2=10^3$. For all three models, we recompute the electron density of lobe cells while keeping the total (simulated) pressure $P_\mathrm{tot}$.

In Fig.~\ref{fig:evolution_absolute}, the initial jet inflates lobes which, after jet shut-down, rise buoyantly in the cluster atmosphere. The high jet power leads to jet velocities approaching the speed of light, where our non-relativistic treatment degrades in accuracy. While this influences the details of the kinematics and shock dissipation \citep{Perucho2017}, our results on bubble morphology and SZ signal are expected to be robust. The jet initially drives a shock wave into the ICM. Generally, the Mach number in the jet direction exceeds that perpendicular to the jet, thereby creating an ellipsoidal shock. The trailing contact discontinuity is clearly visible in the electron number density $n_e$ and temperature $T$ maps. The dimensions of the bubbles and the morphology of the contact discontinuity in the X-ray brightness $l_X$ are in good agreement with observations of MS0735 \citep{Vantyghem2014a}.

Assuming a relativistic power-law distribution of electrons in the bubbles implies a larger contrast of the SZ signal $\delta i_{\nu_0}$. Note, a hypothetical thermal filling of the under-dense bubble with densities as shown in Fig.~\ref{fig:evolution_absolute} implies relativistic temperatures. The high density contrast between bubble and ICM makes the bubble susceptible to the Rayleigh-Taylor instability. Dense gas streams into the lower part of the bubble, generates turbulence and mixes with the bubble gas. This process progressively dilutes the bubble material and transports it into the ICM until complete disruption.

\section{Probing relativistic bubble fillings}
\label{sec:ProbingRelativisticFillings}
The observed decrement in the SZ signal due to different bubble fillings can help to unravel its constituents. In the left panel of Fig.~\ref{fig:radial_sz}, we show the SZ signal across the jet axis for bubbles filled with thermal gas ($kT_e=10\ \mathrm{keV}$) or relativistic gas, which either follows a thermal distribution ($kT_e=500\ \mathrm{keV}$) or a power-law distribution (``relativistic power-law''). The larger SZ contrast for relativistic fillings becomes apparent. 

\cite{Abdulla2018} compared cut-out ellipsoids from a fitted, smooth background cluster with their SZ observations to constrain the bubble filling. Rather than reproducing their analysis, we focus on the consequences for the SZ signal of modelling bubbles as ideal ellipsoids including variations in inclination.

On the right-hand panel in Fig.~\ref{fig:radial_sz}, we contrast the expected SZ signal from our simulated bubble to an ellipsoidal cut-out in our simulation, assuming the unperturbed initial profile. Throughout this Letter, we determine the dimensions of the cut-out ellipsoid and its position from unsharped masked X-ray maps of our simulated bubbles. We assume that the depth of the modelled bubble corresponds to its width. On the right-hand panel in Fig.~\ref{fig:radial_sz}, both bubbles are viewed perpendicular to the jet axis and contain a relativistic power-law filling. The profiles show good agreement at the inner rim of the bubble and start to deviate towards the bubble edges, where we see an enhanced SZ signal in the simulations out to the ellipsoidal shock. This feature corresponds to the increased pressure in the shocked cocoon powered by the outburst and is neglected in the simple cut-out model. This highlights the importance of including the bow shock in the model.

\begin{figure}
\centering
\includegraphics[width=0.465\textwidth]{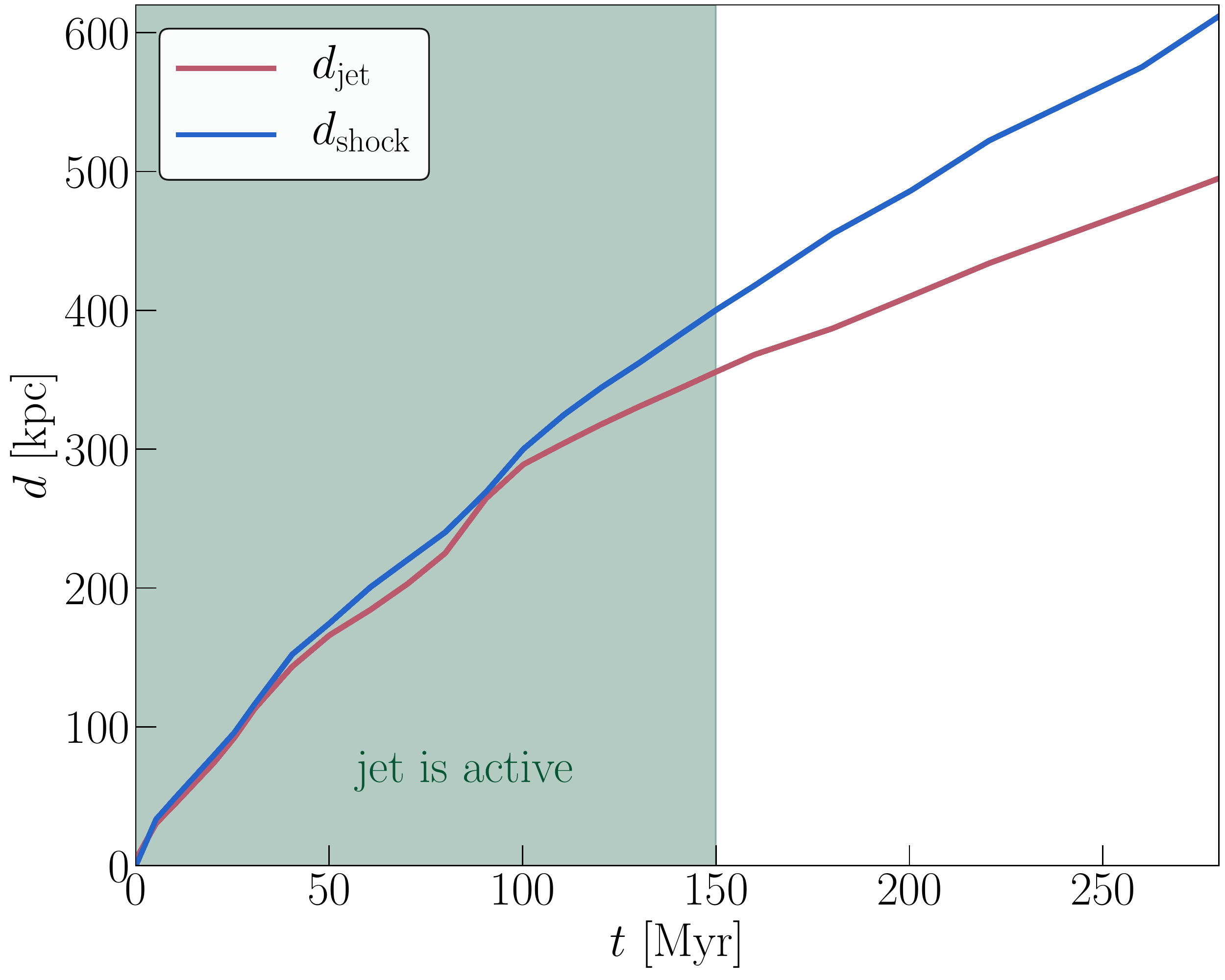}
\caption{Cluster-centric distance of the shock (blue) and head of the jet and bubble, respectively (red) as a function of time. After two thirds of the jet lifetime, the shock detaches and propagates with almost constant velocity $v\approx 1670\,\rmn{km}\,\rmn{s}^{-1}$ while the bubble rises with the slower buoyancy velocity $v\approx 1110\,\rmn{km}\,\rmn{s}^{-1}$, which causes an increasing stand-off distance with time. }
    \label{fig:standoff_dist}
\end{figure}

\section{Degeneracies with jet inclination}
\label{sec:DegeneraciesInclination}
In Fig.~\ref{fig:inclination_comparison}, we study the influence of inclination on the SZ signal. An angle of $\theta=90$ corresponds to a jet axis that is perpendicular to the line of sight. We compare the amplitude of the kinetic SZ effect $w$ (top row of Fig.~\ref{fig:inclination_comparison}) to the thermal SZ effect $y$ (middle row of Fig.~\ref{fig:inclination_comparison}). From left to right, we use later times for larger inclinations to keep the projected distance from the BH to the lower bubble edge approximately constant. We see a strong increase in signal strength of the kinetic SZ effect for larger inclinations. In the wake of the bubble, highly turbulent, dense gas pushes into the region previously occupied by the bubble. This bipolar structure develops significant velocities at high densities that contribute up to 10\% of the thermal effect to the total SZ signal at $\theta=33^\circ$. Future high sensitivity observations of inclined jets should be able to measure this effect. The expected opposite sign of the signal in the two bubbles aids in modelling this effect.

The bottom panels  of Fig.~\ref{fig:inclination_comparison} show the expected SZ signal at $30\ \mathrm{GHz}$ for the differently tilted simulations and the cut-out model, for which we always assume an inclination of $\theta=90^\circ$. In both models, we adopt a power-law distribution of relativistic electrons in the bubbles. The model reproduces the SZ signal at the inner bubble edges well for large inclinations while the agreement becomes worse for smaller inclinations. When viewed face-on ($\theta=90^\circ$), the inclusion of the bow shock in the model is critical to correctly reproduce the SZ signal from the simulation. The bow shock loses its momentum and covers a larger volume as a function of time such that the signal from the shocked region flattens for later times (panels towards the right).

In addition, the SZ contrast due to the relativistically filled bubbles disappears almost entirely for smaller inclinations. Here, lines of sight pass through significantly larger portions of the shocked gas region. This is especially true for the inner part of the bubble where lines of sight intersect the dense central cool core. The effect is amplified for older bubbles through the advanced state of entrainment and  mixing of ICM material with the relativistic bubble contents due to the Rayleigh-Taylor instability (see Section \ref{sec:evolvingSZfromsimulatedbubbles}).

Contamination by the signal from the cluster core appears like a significant
source of uncertainty for the analysis of observations as $\theta$ is
unconstrained a priori. However, X-ray observations can inform simulations about
jet power and bubble/shock ages \citep[e.g.,][]{Diehl2008}. Our simulation shows
that the stand-off distance between bow shock and jet/bubble head is increasing
with time. In particular, entraining the ambient ICM towards the end of the jet
lifetime slows the jet down, see Fig.~\ref{fig:standoff_dist}. Varying the
viewing angle of the jet axis then helps to disentangle projection effects and
potentially constrain inclination.
 
For large inclinations, the projected stand-off distance decreases for decreasing inclinations. However, at small inclinations the outer edge of the ellipsoidally shocked region enlarges the projected bow shock region beyond the projected distance of the upper tip of the bow shock. The ellipticity of the bow shock itself may aid in providing a coarse estimate of the inclination. Even though the details may depend on the concentration of the NFW potential, a spheroidal bow shock should generally favour small angles of inclination. Note that these features have a weak resolution dependence in our simulations, which is studied in \cite{Weinberger2017} and \cite{Ehlert2018}. With high-sensitivity observations the kinetic SZ effect can be identified through its bimodality and used as an additional constraint for the suggested method.

\section{Conclusion}
\label{sec:conclusion}
We show that three-dimensional MHD simulations are instrumental to carefully model the SZ signal of jet-inflated bubbles and conclude:
\begin{itemize}
\item Relativistic bubble fillings imply a large SZ contrast, which is observable in high-resolution SZ observations.
\item The SZ profiles of simulations and the cut-out model show good agreement at the inner rim of the bubble and start to deviate towards the bubble edge because the simplified model fails to account for the shock-enhanced pressure cocoon outside the bubbles.  
\item The match between simulations and model becomes worse when considering small inclinations between line-of-sight and jet axis ($\theta\lesssim45^\circ$). This geometry probes a larger fraction of the shocked ICM, which leads to an increase of thermal SZ signal also towards the inner bubble rim region. Additionally, the light material of the bubble is progressively mixed with denser ICM due to the Rayleigh-Taylor instability, making the bubble indistinguishable from the surrounding ICM. Thus, the SZ signal from older bubbles is reduced in comparison to modelling them with cut-out ellipsoids.
\item At small inclinations, the kinetic SZ effect reaches up to 10\% of the thermal SZ effect. The wake of the bubble causes dense ICM to enter the bubble from below, causing a comparably large kinetic SZ signal. The opposing signs of the signal of inclined bipolar outflows are a smoking-gun signature for identifying this kinetic SZ signal.
\item We propose to constrain the inclination with the stand-off distance between shock and jet/bubble, the elliptical appearance of the bow shock, and (if available) the amplitude of the kinetic SZ effect. To this end, a combination of high-resolution X-ray and SZ observations and full MHD simulations are crucial to break degeneracies due to projection effects.
\end{itemize}
This Letter opens up the possibility to understand biases associated with simplified SZ modeling of AGN bubbles and to finally constrain their contents so that we can observationally identify the physical processes underlying AGN feedback.
\section*{Acknowledgements}
We acknowledge support by the European Research Council under ERC-CoG grant CRAGSMAN-646955.

\bibliographystyle{aasjournal} 

\label{lastpage}
\end{document}